\documentclass[12pt]{article}
\input{amssym.def}
\input{amssym}
\usepackage[dvips]{color}
\usepackage{epsfig}
\usepackage{hyperref}
\normalbaselines
\makeatletter
\@addtoreset{equation}{section}
\makeatother

\def\be{\begin{equation}}
\def\ee{\end{equation}}
\def\ba{\begin{eqnarray}}
\def\ea{\end{eqnarray}}

\newcommand{\rf}[1]{(\ref{#1})}

\def\bra#1{\langle #1|}
\def\ket#1{|#1\rangle}

\begin{document}

\title
{Nonlocal growth processes and conformal invariance}

\author{Francisco C. Alcaraz$^1$   and 
Vladimir Rittenberg$^{2}$
\\[5mm] {\small\it
$^1$Instituto de F\'{\i}sica de S\~{a}o Carlos, Universidade de S\~{a}o Paulo, Caixa Postal 369, }\\
{\small\it 13560-590, S\~{a}o Carlos, SP, Brazil}\\
{\small\it$^{2}$Physikalisches Institut, Universit\"at Bonn,
  Nussallee 12, 53115 Bonn, Germany}}
\date{\today}
\maketitle
\footnotetext[1]{\tt alcaraz@if.sc.usp.br}
\footnotetext[2]{\tt vladimir@th.physik.uni-bonn.de}

\begin{abstract}
Up to now the raise and peel model was the single known example of a 
one-dimensional stochastic process where one can observe conformal 
invariance. The model has  one-parameter. 
 Depending on its value  one has a gapped phase, a 
critical point where one has conformal invariance and a gapless phase with 
changing values of the dynamical critical exponent $z$. In this model, 
adsorption is local but desorption is not. The raise and strip model 
presented here in which desorption is also nonlocal, has the same phase 
diagram. The critical exponents are different as are some physical 
properties of the model. Our study suggest the possible existence of a 
whole class of stochastic models in which one can observe conformal 
invariance.
\end{abstract}

\section{ Introduction} \label{sect1}

There is a long list of papers on one-dimensional interface growth 
models (see \cite{ASE} and \cite{THZ} for reviews). In most of them 
the stochastic processes 
are local. Two typical gapless phases are encountered: the 
Edwards-Wilkinson phase \cite{SEW} where the dynamical critical exponent 
$z = 2$ and the 
Kardar-Parisi-Zhang phase \cite{KPZ} where $z = 3/2$. If in a model one has an 
Edwards-Wilkinson phase by introducing an asymmetry one can get a Kardar-Parisi-Zhang
phase. By introducing "friction"  (some processes oppose growth) one gets a gapped phase (see the flip-flop model 
described below). The interplay of asymmetry and "friction" is nicely displayed 
in \cite{POP}.    

 As far as we know the first example of a nonlocal growth model is the Derrida 
and Vannimenus' study of the interface in weakly disordered systems \cite{DEV}. Much 
later another model with nonlocal rates appeared in the literature, the raise and peel 
model (RPM) \cite{ALC}. This is a one-parameter (denoted by $u$) dependent model which also 
displays three situations, analogously to local models. 
For small values of $u$ one has a 
gapped phase, for $u = 1$ one has a gapless phase with $z = 1$ (not $z = 2$) 
and for $u > 1$   
one has a gapless phase with varying values of $z$ 
(this corresponds to the KPZ phase 
in local models). Dyck (special RSOS) paths describe the interface. 
Adsorption is 
local but desorption is nonlocal. The desorption processes look a bit 
artificial, 
they come from the algebraic background of the model at $u = 1$. 
What is special in 
this model is that for $u = 1$ one can do analytic calculations and 
show that the 
model is conformal invariant. Moreover, the stationary state of the 
model has fascination combinatorial properties \cite{RSOa,RSOb,CAN}.
 Recently, another model (the peak adjusted raise and peel model (PARPM)) was 
introduced \cite{ALCA}. The adsorption and desorption process are like in the RPM but the 
rates depend on the number of peaks in the Dyck paths. This makes the rates 
dependent on the size of the system. A new parameter $p$ was 
introduced such that if $p = 1$, one recovers the RPM at $u = 1$. It was shown that 
conformal invariance is maintained in whole domain of $p$.
 In this paper we present a new model, the raise and strip model (RSM) which is 
again a one-parameter dependent model, with local adsorption and nonlocal 
desorption processes, the rules for the latter being much simpler. 
The configuration 
space is the same as in the RPM. Our aim was to see if the main 
ingredient to get 
conformal invariance is the existence of nonlocal processes of a special kind. 
We were aware that the price to pay is loosing integrability 
and that one had to use 
Monte Carlo simulations on large lattices to get results. We have indeed 
observed that 
the RSM has a phase diagram similar to the one seen in the RPM. 
At the conformal 
invariant point, the critical exponents are different. 
Since other models having the 
same structure as the RSM can easily be defined, it is plausible to 
assume that 
there is a whole class of models with conformal invariance which should 
 be studied.

 The paper is organized as follows. In Section 2 we defined the observables for 
models defined on Dyck paths. 
These observables are used in the description of the 
properties of the models.
 
The flip-flop, the raise and peel and raise and strip models are defined in Section 3.
The flip-flop model is the local version of the RSM, it was studied in order to see 
the effect of nonlocality introduced in the RSM.

 The flip-flop model is presented in Section 4. One shows, using known results from 
combinatorics and Monte Carlo simulations that one has a critical point  with a 
dynamic critical exponent $z = 2$ which separates two gapped phases. 
        
 The raise and peel model is shortly reviewed in Section 5.

 The main results of our research are given in 
Section 6 in 
which we present not 
only the properties of the stationary states of the RSM but also the time dependent phenomena. 
An interesting new phenomenon occurs if the parameter $u$ is larger the 
$u_{c}$. 
The system stays gapless with varying values of $z$ (like in the RPM) but 
unlike the RPM 
where in the stationary state, the average height 
increases logarithmically with 
the size of the system, in the RSM, the average height profile 
is a triangle with a 
height of the order of the system size.

For completeness, a variant of the flip-flop model in which the 
configuration space is changed, is presented in the Appendix.

 Our conclusions can be found in Section 7

\section{ 
 Observables for models defined on Dyck paths}

 We consider an open one-dimensional system with $L + 1$
sites ($L$ even). A
Dyck path is a special restricted solid-on-solid (RSOS) configuration  defined as
follows. We attach to each site $i$  integer heights $h_i$
 which
 obey RSOS rules:
\be \label{e1}
 h_{i+1} - h_i =\pm 1, \quad   (i = 0,1,\ldots,L-1),
\ee
with the constraints:
\be \label{e2a}
h_0 = h_L = 0,
\ee
\be \label{e2b}
 h_i \geq 0, \quad (i=0,1,\ldots,L).
\ee
There are
\be \label{e3}
Z (L) = L!/(L/2)!(L/2 + 1)!
\ee
configurations of this kind.
\begin{figure}
\centering
\includegraphics[angle=0,width=0.5\textwidth] {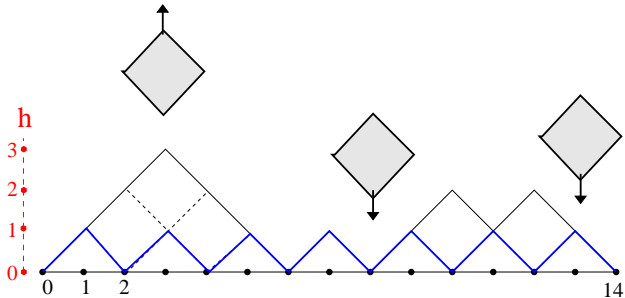}
\caption{
 An example of a Dyck path for $L = 14$.  There are four contact
points, three clusters,  four peaks and three valleys. The substrate profile is shown in blue.}
\label{prof1}
\end{figure}

 A Dyck path can be seen as an interface separating a film of tilted 
tiles deposited on a substrate, from a rarefied gas of tiles (see Fig.~\ref{prof1}). 
The substrate corresponds to the special Dyck path defined as $h_{2k} = 
0$, $h_{2k+ 1} = 1$ ($k=0,1,\ldots,L/2-1$).

  If $h_j = 0$ at the site $j$, one has a {\it contact point} (there are 
four contact points in Fig.~\ref{prof1}). Between two consecutive contact points one
has a {\it cluster} (there are three clusters in Fig.~\ref{prof1}). The slope at the
site $i$ is $s(i) = (h_{i+1} - h_{i-1})/2$. If $s_i = 0$
($h_i > h_{i-1}$) one has a {\it peak}. If $s_i = 0$ ($h_i < h_{i-1}$)
one has a {\it valley} (there are four peaks and three valleys in Fig.~\ref{prof1}). 

 It is useful to use the known mapping between an RSOS path and the 
configuration space of a one-dimensional hopping model with 
exclusion.
 To an upwards step $h_i - h_{i-1} > 0$ one associates a particle, 
to an downwards step $h_i - h_{i-1}<0$ one associates a vacancy. The 
constraints \rf{e2a}-\rf{e2b} are translated into two conditions on the 
particles-vacancies configurations. The first constraint \rf{e2a}
 implies that 
the number of particles is equal to the number of vacancies. The second
constraint \rf{e2b} is  obviously nonlocal: the number of particles on the left
side of any bond has to be larger or equal to the number of vacancies 
on the left side of the same bond. For example, the configuration 
$XXOXOO$ is acceptable but not $OOXOXX$ ($X$ is a particle, $O$ is a vacancy).
The Dyck path shown in Fig.~\ref{prof1} corresponds to the configuration 
$XXXOOOXOXXOXOO$.
  
 The raise and fall stochastic models to be described below give 
 the probabilities of the various Dyck paths and one is interested in 
average values of observables. 

 One  obvious observable is $h(i,L,t)$ which is the average height at the 
site $i$ for a system of size $L$ and  time $t$. The average 
density of contact points which is a function of the site $i$, the 
size $L$ of the system and $t$, will be denoted by $g(i,L,t)$. The
average density of clusters equal to the average number of clusters 
divided by $L$, will be denoted by $K(L,t)$. When one considers the 
stationary states of the models, the time dependence will be dropped 
in the notation. For example, $g(i,L,t)$ will become $g(i,L)$.
It turns out that the average density of peaks and valleys $\tau(L,t)$, equal
to the average total number of peaks and valleys divided by $L$, plays
an 
important role in studying the properties of the models.

\section{ Raise and fall models}

 We present three stochastic models defined in the configuration space of
Dyck paths. 
In one of the models (the flip-flop model) the processes are local, in the 
other two (the raise and peel and raise and strip models), the adsorption 
processes are nonlocal. 
As we are 
going to see it is the nonlocality of the rates which is relevant in getting
 new physics.

The models depend on one parameter $u$ which is the ratio of adsorption
and desorption rates.
 One uses sequential updating. At each time step, with a probability $1/(L-1)$ a
tile hits the Dyck path at a site $i = 1, 2 ,\ldots, L-1$. The effects of the hits
are different in the three models.
\begin{figure}
\centering
\includegraphics[angle=0,width=0.5\textwidth] {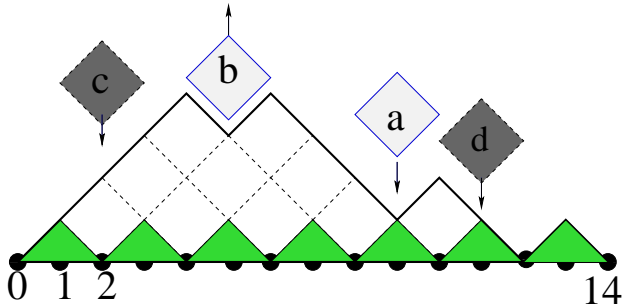}
\caption{
The flip-flop model. When a tile hits the interface, the 
following processes occur: The tiles $c$ and $d$ are reflected, tile $b$ 
triggers a local desorption and the tile $a$ which hits a valley, is 
adsorbed.}
\label{prof2}
\end{figure}

 The {\it flip-flop model} (FFM) is an extension of 
the freely jointed chain model
of a random coil polymer \cite{VER}. In this model both the adsorption and
desorption processes are local. The nonlocality comes only from the
constraint \rf{e2b} which defines the Dyck paths and, unlike the two other
models, not from the rates. The model is defined by the following rules:

 If a tile hits a valley ($s_i = 0$ and $h_i < h_{i-1}$), with a rate $u$ it sticks
to the site and the valley becomes a peak (see tile $a$ in Fig.~\ref{prof2}). If a tile hits a
 peak ($s_i = 0$ and $h_i > h_{i-1}> 0$), with a rate equal to one,
 the tile at the peak gets desorbed and the peak becomes a valley (see tile $b$ 
in Fig.~\ref{prof2}). If the tile hits a
site $i$ and $s_i \neq 0$, the tile is reflected with no changes in the 
profile (see tiles $c$ ans $d$ in Fig.~\ref{prof2}).
In this model the rates are local. In the raise and strip model 
to be described below 
the desorption rates are nonlocal and it is interesting therefore to compare
 the 
models.

 The {\it raise and peel model} (RPM) was intensively studied \cite{ALC}. We
present it here in order to clarify the effects of nonlocality on the
physics of the models.
\begin{figure}
\centering
\includegraphics[angle=0,width=0.5\textwidth] {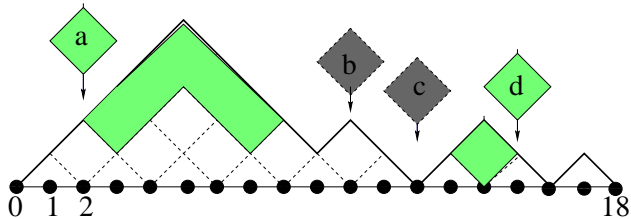}
\caption{
The raise and peel model. When tile $a$ hits the interface it triggers a
nonlocal desorption. The desorption can be local only if the tile
hits the substrate (tile $d$). If the tile hits a peak, it is reflected 
(tile $b$).
If a tile hits a valley, like in the flip-flop model, the tile is adsorbed
(tile $c$).}
\label{prof3}
\end{figure}

 The main merit of the RPM is that for $u = 1$, it is integrable and
conformal invariant. The essential features of the model are local
adsorption and nonlocal desorption processes of a particular kind. The
rules for the latter come from the algebraic structure (the Temperley-Lieb
algebra) behind the model. These rules  do not easily generalize for other
configuration spaces where we expect to be able, based on algebraic
considerations (using Hecke algebras), to define stochastic models which are
also conformal invariant. An example is the case of restricted Motzkin
paths \cite{DIF}. The RPM is defined by the following rules:

Depending on the slope $s_i=(h_{i+1}-h_{i-1})/2$
at the site $i$, the following processes can occur:

\noindent 1) $s_i = 0$ and $h_i < h_{i-1}$ (tile $c$ in Fig.~\ref{prof3}). The tile hits a local minimum and   with a rate $u$ is adsorbed ($h_i\rightarrow  h_i + 2$).

\noindent 2) $s_i = 0$ and $h_i > h_{i-1}$ (tile $b$ in Fig~\ref{prof3}). The tile hits a peak and is reflected.

\noindent 3) $s_i = 1$ (tile $a$ in fig.~\ref{prof3}). With a rate one the tile is reflected after triggering the desorption ($h_j \rightarrow h_j-2$) of a layer of $b-1$ tiles from the segment $\{j=i+1,\ldots,i+b-1\}$ where $h_j>h_i=h_{i+b}$.

\noindent  4) $s_i = -1$ (tile $d$ in Fig.~\ref{prof3}).  With a rate one, the tile is reflected after
triggering the desorption ($h_j \rightarrow h_j-2$) of a layer of
$b-1$ tiles belonging to the segment $\{j=i-b+1,\ldots,i-1\}$
where $h_j>h_i=h_{i-b}$.

\begin{figure}
\centering
\includegraphics[angle=0,width=0.5\textwidth] {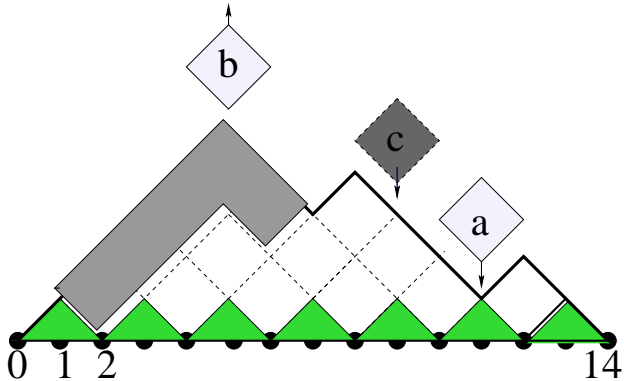}
\caption{
The raise and strip model. When a tile hits a peak it triggers a nonlocal 
desorption process (tile $b$), the tile is reflected otherwise (tile $c$) 
unless it hits a valley when it gets adsorbed (tile $a$).}
\label{prof4}
\end{figure}

 The {\it raise and strip model} (RSM) was conceived in order to keep the
main features of the RPM (local adsorption and nonlocal desorption) using
simpler rules and to see if one can recover conformal invariance. The
price to pay is of course, lack of integrability and other magic
properties  of the RPM. Keeping  in mind that in a Dyck path each valley is
followed by a peak and that a peak is surrounded by two consecutive valleys,
the RSM is defined by the following rules:

 If a tile hits a valley ($s_i = 0$ and  $h_i < h_{i-1}$), with a rate $u$
 it sticks
to the site and the valley becomes a peak (see tile $a$ of Fig.~\ref{prof4}). If a tile hits a
peak ($s_i = 0$ and  $h_i > h_{i-1}> 0$), surrounded by two valleys at
the sites $k$ and $l$  ($k < i < l$), with a rate equal to one, a layer of tiles
between the two consecutive valleys is desorbed (see tile $b$ of Fig.~\ref{prof4}). One has $h_j \rightarrow  h_j - 2$
($j = k + 1, k + 2, \ldots ,l - 1$). If the tile hits a site $i$ and 
$s_i \neq 0$,
like in the FFM but unlike the RPM, the tile is reflected (see tile $c$  of 
Fig.~\ref{prof4}).

 We notice that the adsorption process is common to all three models. The
differences are in the desorption processes. They are similar in
the RSM and RPM: in general a layer of tiles evaporates and not a single tile. 
This
is why one expects to find in the RSM model a value of the parameter $u$ for
which one could see conformal invariance like in the RPM.

 In the FFM and RSM the desorption takes place when a tile hits a peak and is
reflected when the tile hits a site with no valleys or peaks. In the first
model desorption is local but is nonlocal in the RSM. The number of active
sites for desorption is equal to the number of active sites for adsorption.
In the RPM this is not anymore the case, the number of active sites for
desorption being the sites with $s_i \neq 0$.

 The differences between the three models mentioned above will determine
major differences in their physical properties which are going to be
discussed in the next sections.


In order to study the continuous time evolution in the three models described 
above, one uses the master equation which can be interpreted as an imaginary time 
Schroedinger equation. If the system is composed by the states $a = 1,2,..., Z(L)$, 
the probabilities $P_a(t)$ are the solutions of the equation:
\begin{equation}\label{e4}
\frac{d}{dt} P_a(t) = -\sum_b H_{a,b} P_b(t).
\end{equation}
 The Hamiltonian $H$ is an $Z(L)\times Z(L)$ intensity matrix: $H_{a,b}$
($a\neq b$) is non positive
 and $\sum_a H_{a,b} = 0$. $-H_{a,b}$ ($a\neq b$) is the rate for the transition
$\ket b \rightarrow \ket a$. The ground-state wavefunction of the system $\ket0$, $H \ket0 = 0$, gives
the probabilities in the stationary state:
\begin{equation} \label{e4p}
\ket0 = \sum_a P_a \ket a,\;\;\;\;\;\; P_a = \lim_{t \to \infty}  P_a(t). 
\end{equation}

 In order to go from the discrete time description of the stochastic model
to the continuous time limit, we take $\Delta t = 1/(L-1)$ and
\be \label{e5p}
H_{ac}= - r_{ac}  \quad (c\neq a),
\ee
where $r_{ac}$ are the rates described above for each of the three models.

\section{ The flip-flop model}

 We will first discuss the stationary state of the model using some known
results in combinatorics. We will discover in this way that we have three
phases. One for $u < 1$, one for $u = 1$ and another one for $u > 1$. What should we
expect to find?

\begin{figure}
\centering
\includegraphics[angle=0,width=0.5\textwidth] {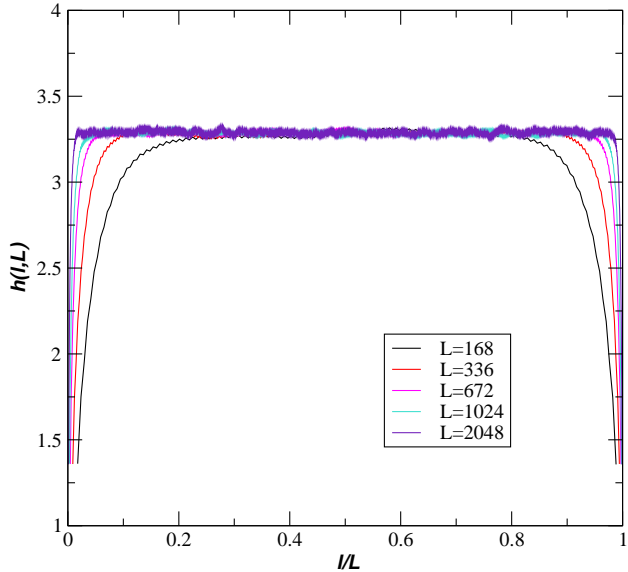}
\caption{
 The average height at a site $l$ for different values of $L$ in the 
stationary state of the flip-flop model for $u=0.95$ 
(under the critical point). The 
lattice sizes are: $L = 168, 336, 672, 
 1024$ and 2048.}
\label{avdens}
\end{figure}

 If $u < 1$, evaporation takes over deposition of tiles. One should therefore
find in the stationary states, in the thermodynamical limit, a finite average
height.  This assumption is confirmed by Monte Carlo simulations at $u = 0.95$
for various lattice sizes. The results are shown in Fig.~\ref{avdens}. One can observe
a constant (site independent) average height. The finiteness of the
heights give finite values for the estimators of shared information \cite{RAA} and
therefore we are in a gapped phase.

 For $u = 1$, all the Dyck paths have the same probability and therefore can
be seen as the paths of a restricted random walker who starts at the origin and
returns after $L$ steps. Since it is a random walker, this implies that one is in a
gapless phase with a dynamic critical exponent $z = 2$ \cite{VER}.

 For $u > 1$, Dyck paths with large heights are preferred and one can expect a
growing interface with $z = 3/2$ corresponding to the KPZ universality class \cite{KPZ}.
In this consideration we didn't take into account the constraints. We are
going to see that our guess is not necessarily correct. 
 We proceed by presenting the case $u>1$ in detail.

a) {\it The stationary state.}

\begin{figure}
\centering
\includegraphics[angle=0,width=0.5\textwidth] {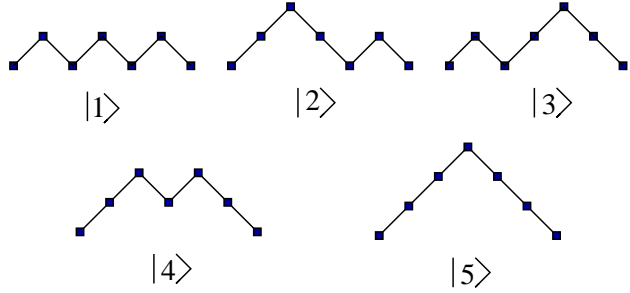}
\caption{The Dyck path  
  of 
the $L=6$ lattice ($(L+1)$ points).}
\label{rsos6}
\end{figure}

It is instructive to take $L = 6$ and write the Hamiltonian \rf{e5p} in the
vector space of the five Dyck paths ($Z(6) = 5$). The five configurations are
shown in Fig.~\ref{rsos6}.

The configuration 
$\ket{1}$ corresponds to the 
substrate. Configuration $\ket{5}$ in which $h(L/2,L) = L/2$ 
is  the 
{\it pyramid } configuration.

The matrix  elements of the Hamiltonian are:
\ba
 \label{4.1}
&& H =-
\left( \begin{array}{c|rrrrr}
 & \ket{1} & \ket{2} & \ket{3} & \ket{4} &\ket{5} \\ \hline
\bra{1} &   2u &    -1 &   -1  &  0 & 0 \\
\bra{2} &   -u &  1+u &   0 & -1 &      0 \\
\bra{3} &   -u  &   0   & 1+u &  -1 &    0 \\
\bra{4} & 0  &  -u  &  -u &   2+u  &   -1  \\
\bra{5} &     0 &  0 &  0 & -u &   1
\end{array} \right) .
\ea

 The wavefunction corresponding to the eigenvalue zero is:

\be \label{4.2}
\ket{1} + u\ket{2} + u\ket{3} + u^2\ket{4} +u^3\ket{5}.
\ee

 Notice that in the stationary state wavefunction each configuration gets
as a coefficient a monomial in $u$ with an exponent equal to the number of tiles
on the top of the substrate. We have chosen the coefficient of the substrate (no
tiles) to be equal to one. For large values of $u$, the configuration $\ket{5}$
corresponding to the {\it pyramid} is preferred, for small values of $u$, the
configuration $\ket{1}$, which corresponds to the substrate has the largest probability. The normalization factor
\be \label{4.3}
Z_6(u) = 1+2u+u^2+u^3
\ee
is equal to the generating function for the number of tiles.
 One can easily check that this observation is valid for any number of
sites and that the normalization factor $Z_L(u)$ is the generating function for
the number of tiles for any lattice size $L$. Once this point is made, one
can use results from combinatorics to get the phase diagram of the model.

 It turns out \cite{FUH} that $Z_L(u)$ is related to the Carlitz $q$-Catalan numbers
$C_n(q)$($L = 2n$). The latter are defined by the recurrence relations:
\be \label{4.4}
C_{n+1}(q) = \sum_{k=0}^n C_k C_{n-k}q^{(k+1)(n-k)}, \quad (C_0=1).
\ee
We give the first ones:
\ba \label{4.5}
&&C_1=1, \quad C_2=1+q, \quad C_3 = 1+q +2q^2+q^3, \nonumber \\
&& C_4=1+q+2q^2+3q^3+3q^4+3q^5+q^6.
\ea

 One can define another deformation of the Catalan numbers
\be \label{4.6}
\tilde{C}_n(u) = u^{\frac{n(n-1)}{2}}C_n(u^{-1}),
\ee
which are the solutions of the recurrence relations
\be \label{4.7}
\tilde{C}_{n+1}(u) = \sum_{k=0}^n u^k \tilde{C}_k(u)\tilde{C}_{n-k}(u), \quad (\tilde{C}_0=1).
\ee

 One has
\be \label{4.8}
Z_L(u) = \tilde{C}_n(u) \quad (L=2n).
\ee

 The average number of tiles is
\be \label{4.9}
N_L(u) = u \frac{d}{du} \ln{Z_L(u)}.
\ee

 There is no known explicit expression for $Z_L(u)$ for finite values of $L$
and its  asymptotics is known only  for $u > 1$ \cite{FUH}:
\be \label{4.10}
\lim_{n \to \infty} Z_{2n}(u) = u^{\frac{n(n-1)}{2}} /\phi(u^{-1}),
\ee
where $\phi(q)$ is the Euler function
\be \label{4.11}
\phi(q) = \prod_{n=1}^{\infty}(1-q^n).
\ee

 Using \rf{4.9} and \rf{4.10} one can compute the average number of tiles in the
large $L$ limit. One gets
\be \label{4.12}
\lim_{L\to\infty}N_L(u) = \frac{n(n-1)}{2} + C(u),
\ee
where the $L$-independent term $C(u)$ is:
\be \label{4.13}
C(u) = - \sum_{k=1}^{\infty} \frac{k}{u^k -1}.
\ee

\begin{figure}[t]
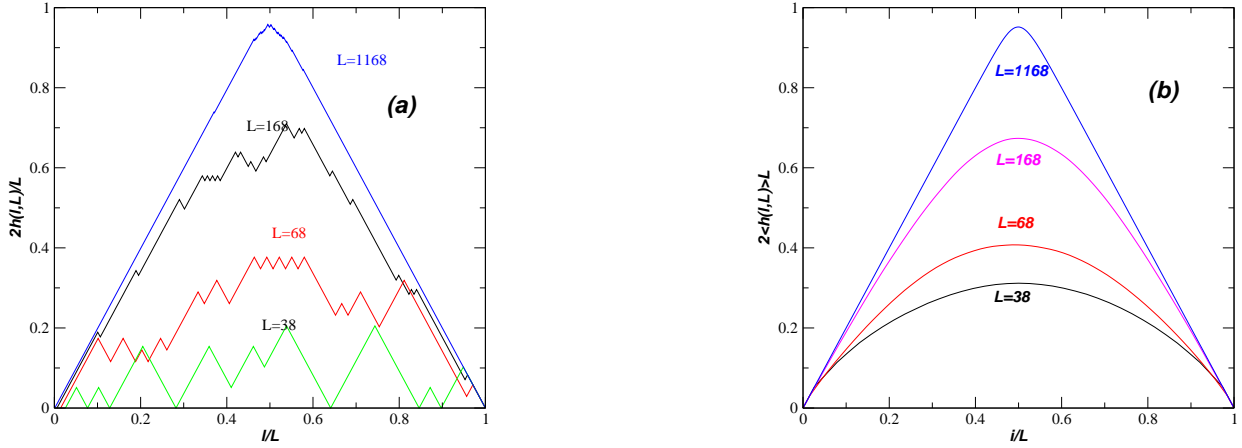

\centering{ \includegraphics [angle=0,scale=0.46]
{typa.eps}\hspace{-3.0cm}\hfill\includegraphics
[angle=0,scale=0.46] {typb.eps} } \caption{
a) Typical configurations of the heights $h(l,L)$ (multiplied by $2/L$) in
the flip-flop model for $u = 1.05$ (over the critical point) and
different
 system sizes ($L = 38$, $68$, $168$ and $1168$). For the pyramid
configuration, $2h(L/2)/L = 1$.
b) Average values, considering $10^5$ configurations for the same systems in (a).}
\label{typa}\label{typb}
 \end{figure}

 It follows that the average number of tiles is equal to those in the
pyramid $n(n-1)/2$ for any $u$! As a consequence, the dominant configurations
are those close to the full "pyramid" for any $u > 1$.
 Monte Carlo simulations for finite values of $L$ confirm this result. Taking
$u = 1.05$ we show in Fig.~\ref{typa}a some typical heights profiles for several
lattice sizes and in Fig.~\ref{typb}b the average values of the heights. One sees
that with increasing values of $L$, one reaches the "pyramid" configuration.

 In the Appendix we consider a model in which the adsorption and 
desorption processes are the same as in the flip-flop model but the 
configuration space is not anymore Dyck paths, the constraint \rf{e2b} being 
relaxed. In the large $L$ limit, the physics is the same as the one observed 
in the flip-flop model for $u > 1$.

 One expects that the whole $u > 1$ domain to be gapped and that for any initial
condition, the systems  evolve fast to reach the configurations closed
to the full pyramid. We are going to show below that this is indeed the
case.

%

 One can understand this result in a different way by using the {\it Dyck paths $\rightarrow$ particle}
 mapping. The desorption process {\it peak $\rightarrow$ valley} with a rate one is mapped
in the hopping of a particle to the right:
\ba 
&& X + 0 \rightarrow 0 + X, \quad {\mbox {rate}} \quad p_R = 1.
\ea
The adsorption process {\it valley $\rightarrow$ peak} with a rate $u$ is mapped in the
hopping to the left:
\ba 
 0 + X \rightarrow X + 0, \quad {\mbox {rate}} \quad p_L = u.
\ea
 We have to take the two constraints into account. The system is confined
to the size $L$ and the number of particles to the left of each bond has to
be larger than the number of vacancies. One can consider ASEP with
boundaries \cite{ASE} and try to mimic the constraint by
injecting with a large rate $\alpha$ particles on the first site and removing
particles with a rate $\beta$ = $\alpha$ (the density of particles has to be
equal to the number of vacancies). This picture should be correct if 
$p_L > p_R$
($u > 1$) when one has a reverse bias. The partition function was computed
\cite{BMF} and one obtains for large systems, a factor $u^{n^2/16}$, similar to
Eq.~\rf{4.10}. In both cases the exponent is an area and not a length. The
current also vanishes in this limit. There is no current in our model.

 This mapping gives absurd results if $u < 1$ ($p_R > p_L$). Since we have
equal densities, in ASEP with boundaries we are in the maximum current
phase which  is gapless. In our model there is no current, there are no density
fluctuations  and one is gapped.

 To sum up, the study of the stationary state suggests the following phase
diagram for the flip-flop model. It has a gapped phase for $u < 1$, is gapless
at $u = 1$ with $z = 2$ and gapped for $u > 1$. We now show that the time
evolution results confirm this picture.

b) {\it Gaps and the dynamical critical exponent} $z$

\begin{figure}
\centering
\includegraphics[angle=0,width=0.5\textwidth] {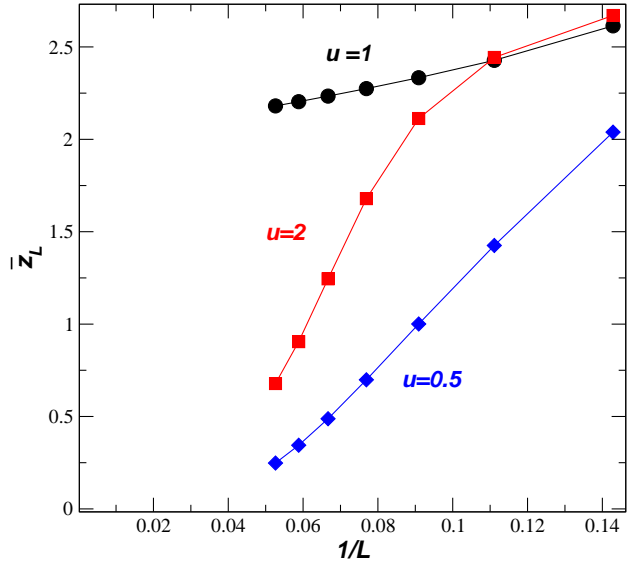}
\caption{
Estimates of the dynamic critical exponent $z$ using \rf{4.15}, as a function of
$1/L$, in the three phases.}
\label{gaps1}
\end{figure}

 In order to see which phases are gapped or gapless, we have diagonalized
numerically (up to $L = 18$) the Hamiltonian \rf{e5p} for the FFM. Since the lowest energy
is zero (the Hamiltonian describes a stochastic process), the energy gap is
given by the first excited state $E_1(L)$. One expects
\be \label{4.14}
E_1(L) = a_1L^{-z},
\ee
where $z$ is the dynamical critical exponent and $a_1$ a constant. 
A value $z = 0$
implies that the system is gapped. In Fig.~\ref{gaps1} we show  the estimates for $z$ 
defined as
\ba \label{4.15}
&&\bar{z}_L = \ln\left(\frac{E_1(L)}{E_1(L-2)}\right)/\ln\left(\frac{L-1}{L}\right),
\ea
up to $L = 18$, in the three phases. One sees that $\bar{z}_L$, as expected, 
 goes 
towards 
the value 2 for $u = 1$ and to zero in the  other two phases.
 To sum up, the phase diagram of the flip-flop model is as follows:
 \be \label{4.16}      
  u < 1\quad \mbox{ gapped,} \quad u = 1 \quad \mbox{gapless} \quad 
(z = 2),\quad  u > 1 \quad \mbox{gapped}.    
\ee

 A version of the flip-flop model in which the rates are the same,  but with a
 configuration 
space without the constraint \rf{e2b}, is presented in Appendix A.

\section{ The raise and peel model - some results}

As a result of the changes occurring in the adsorption process 
which is nonlocal as compared
to the FFM where it is local, the physical properties of the phases in the same phase
diagram also change. Like in the FFM, the $u < 1$ domain is gapped. Instead of a
$z = 2$ phase transition, at $u = 1$ one has a $z = 1$
phase transition with a space-time symmetry (conformal invariance). For $u > 1$,
the system stays gapless with varying dynamic critical exponents $z$ 
 ($z$ decreases when $u$ increases) and is
not gapped like in the FFM.

 Since in the RSM we will find again a $z = 1$ phase transition with
conformal invariance, we will sum up the main results obtained in the RPM
at $u = 1$ \cite{ALC} in order to compare them in the next sections with those
which will be seen in the RSM.

 First we discuss the stationary state.
 In the finite-size scaling limit $l, L >> 1$, $l/L$ fixed, the average height
at a distance $l$ from the origin, for a system of size $L$, has the
expression:
\be \label{5.1}
h(l,L) = \frac{\sqrt{3}}{2\pi}\ln L_c,
\ee
where
\ba \label{5.2}
&&L_c = \frac{L}{\pi}\sin\left(\frac{\pi l}{L}\right).
\ea

 We would like to mention that for $u > 1$, similar to \rf{5.1}, $h(l,L)$ 
shows a logarithmic increase with the size of the system $L$ but with a different dependence on $l/L$.

The density of contact points in the same limit, has the expression
\be \label{5.3}
g(l,L) = \frac{\alpha}{L_c^{1/3}}
\ee
with
\be \label{5.4}
\alpha = -\frac{\sqrt{3}\Gamma(-1/6)}{6\pi^{5/6}}= 0.753149\ldots.
\ee
 These expressions are exact. Notice the logarithmic growth of the
interface and that the lengths dependence is all in $L_c$. The latter is a
consequence of conformal invariance.

 For $u < 1$, in the vicinity of the critical point $u = 1$, and large values of $L$, the 
average density of clusters $K(u)$ is $L$ independent and vanishes like

\be \label{5.5}
K(u) = 0.596 (1-u)^{0.78}.
\ee
This result is new. It was obtained using Monte Carlo simulations. This 
result is a 
surprise. The scaling dimensions in the model are $1/3$ and $1$ \cite{RS} 
and no combination of 
them can give a number close to the exponent $0.78$.

 The average density of peaks and valleys, in the large $L$ limit, is
\be \label{5.6}
\tau = \frac{3}{4}.
\ee

 The physical meaning of \rf{5.6} is the following one. Adsorption takes place
only on valleys which occupy a 3/8th fraction of the number of sites,
whereas desorption takes place on a 2/8th fraction of the number of sites.
Since in a desorption process one loses more tiles than one gains in an
adsorption process, one needs more sites with valleys than sites where
desorption can take place. In the RPM peaks are not active. This picture
will change in the RSM in which sites with both valleys and peaks are
active.

 The spectrum of the Hamiltonian \rf{e5p}, in the finite-size scaling limit, 
is given by
\be \label{5.7}
\lim_{L\to \infty}E_i(L) = \frac{\pi v_s}{L}\Delta_i, \quad  i=0,1,2,\ldots,
\ee
where $E_0 = 0$, $\Delta_i$ are the scaling dimensions and the sound velocity
$v_s$ has the value
\be \label{5.8}
v_s = \frac{3\sqrt{3}}{2}.
\ee
 The scaling dimensions $\Delta_i$ and their degeneracies ($d_i$) can be
obtained from the partition function \cite{RS}
\be \label{5.9}
Z(q) = \sum_{i=0}q^{\Delta_i} = (1-q) \prod_{n=1}^{\infty}(1-q^n)^{-1}.
\ee
 We give the first values of $\Delta_i$ and $d_i$:
\be \label{5.10}
 \Delta = 0(1), 2(1), 3(1), 4(2),\ldots .  
\ee

 As we are going to see in the next Section, the RSM has a phase diagram
similar to the RPM. For $u < u_{c}$ the system is gapped, at $u = u_{c}$ it is
conformal invariant and gapless for $u > u_{c}$
 with varying dynamical critical exponent $z$. The value of $u_{c}$ 
 being not
equal to one anymore. Which properties described above should one expect at
$u_{c}$?

 In the stationary state, the $l$ and $L$ dependence in the finite-size
limit of the average density of contact points and heights should be again
through the function $L_c$ given by \rf{5.2}. The exponents might be different.
How about the spectrum of the Hamiltonian \rf{e5p}? Since the scaling dimension 
$\Delta_1=2$ 
corresponds  to the energy-momentum tensor, it should be not degenerate.
The other values of $\Delta_i$ should be present but they might have other
degeneracies.

 Using $E_1(L)$ and \rf{5.7} with $\Delta_1 = 2$, one can determine $v_s$. The
ratios $E_i(L)/E_1(L)$ should be equal to $n/L$ ($n = 3,4,\ldots$) for 
large $L$.

\section{ The raise and strip model}

We are going to show that this model has a phase diagram similar to those 
found in   the
previous two models. A gapped phase for $u < u_c$, a critical point at 
 $u_c$
and a new gapless phase for $u > u_c$. In establishing the phase diagram one
encounters a new problem: $u_c$ is not known exactly and one has to take
care of possible cross-over effects.

\begin{figure}
\centering
\includegraphics[angle=0,width=0.5\textwidth] {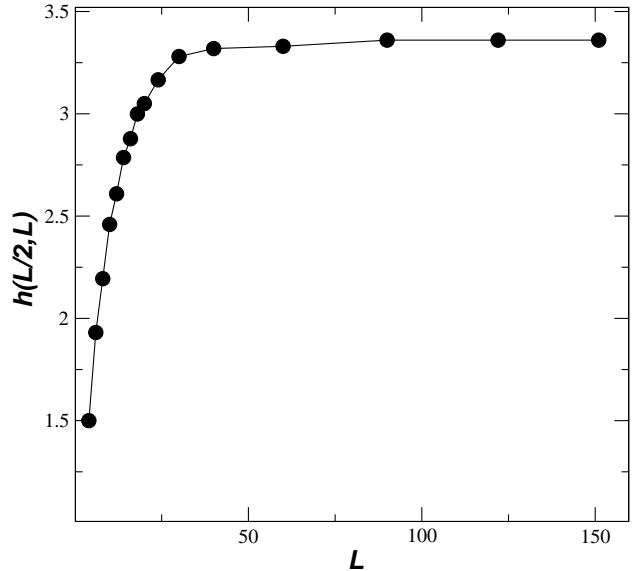}
\caption
{ The RSM. The height in the middle of the system $h(L/2,L)$ for $u = 3$ 
as a function 
of $L$. The results were obtained by averaging $10^8$ samples. The estimated 
errors are smaller than the symbols representing the data.}
\label{rsmhei1}
\end{figure}

 We have analyzed, using Monte Carlo simulations, the heights profiles for
different values of $u$ and lattice sizes and found that up to $u$
around 4.5 they are flat  and the average height is $L$ independent for large values of $L$,  suggesting a 
gapped phase. In Fig.~\ref{rsmhei1}    we show for $u = 3$, the $L$ dependence of $h(L/2,L)$,     
the average height in 
the middle of the system. One sees that its value saturates fast 
with increasing values of $L$.

\begin{figure}
\centering
\includegraphics[angle=0,width=0.5\textwidth] {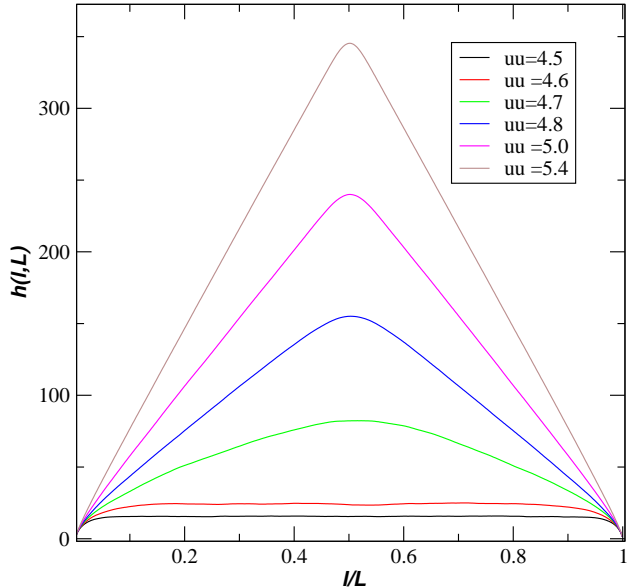}
\caption{
 Average height $h(l,L)$ as a function of $l$ in the RSM 
for various values of $u$:
4.5, 4.6, 4,7, 4.8, 5.0 and 5.4. The size of the system is $L = 2048$.
The average is taken over $10^8$ independent samples.}
\label{rsmhei2}
\end{figure}

 The values of the
average heights are larger than those observed in the RPM. For larger
values of $u$ there is a sharp increase of the height in the middle of
the system. In Fig.~\ref{rsmhei2} we show the heights profiles in the domain
$4.5 \leq u \leq 5.4$ for $L = 2048$. 
One sees that for $u = 5.4$ one sees almost
a triangle with a
rounded tip. A similar triangle was seen in the flip-flop model in the
$u > 1$ domain. There is however a major difference between the triangles
observed in the two models. Whereas in the FFM the triangle coincides with
the pyramid  configuration for which the height in the middle is equal to 
$L/2$, for $u 
= 5.4$ in 
the RSM the average height in the middle is around 30\% of $L/2 = 1024$. 
If one considers larger values of $u$ the height of the triangle increases but
never reaches the pyramid's height. These observations suggest the
existence of a new phase in which the  system grows linearly with
the
size of the system. We will denote this phase by LG (linear growth). We
will learn more about it below.

 A closer inspection of the interval $4.6\leq u \leq 4.7$  suggests a phase
transition between the  gapped phase and the LG phase. In this
interval the profiles
are not sharp at the boundaries as in the gapped phase (one has an
exponential fall-off in this case) and the height at the profile is not a
triangle. One expects therefore $u_c$ to be in this interval.

 We proceed now to a detailed analysis of the model. There is not
much to say about the gapped phase. As we are going to show below, a
good estimate for $u_c$ is $u_c = 4.685$. We have measured using
Monte Carlo simulations the density of clusters in the gapped
phase and found for $u$ close to $u_c$ 
\be \label{6.1}
K(u) = 0.035 (4.685 - u)^{1.733}.  
\ee
 This expression is quite different of the one observed for the RPM
(5.5). There are fewer contact points in the gapped phase in the RSM
than in the RPM. This is consistent with the observation that the
 heights in the gapped phase are higher in the RSM than in the RPM.

 We first discuss the $u_c$ physics and show that we have conformal
invariance for this value of $u$. We present the results for the
stationary state and time dependent observables. The LG phase is
going to be discussed afterwards.

a) {\it The stationary state at $u_c$. Conformal invariance.}

\begin{figure}
\centering
\includegraphics[angle=0,width=0.5\textwidth] {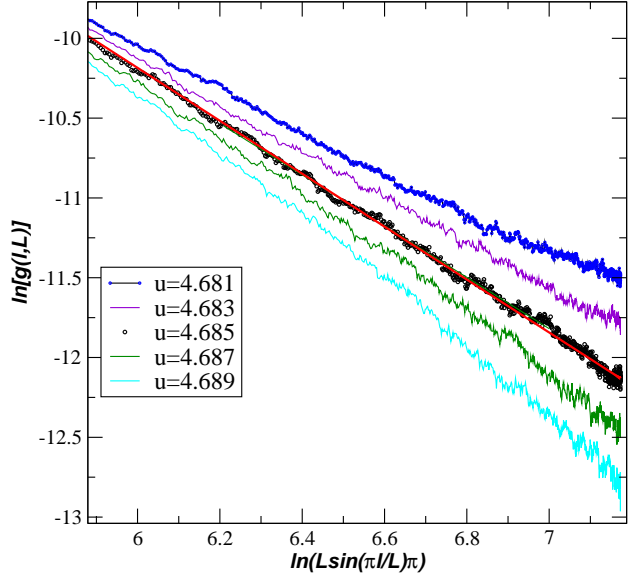}
\caption{
 $\ln{[g(l,L)]}$ in the RSM as a  function of $L_c$ for $L= 4096$ and for the values of: $u =
4.681, 4.682,...,4.689$. 
The results are obtained by averaging over $10^9$ Monte Carlo steps.
The fitting curve at $u=4.685$ is $y=-0.234 -1.65 x$.}
\label{confinv1}
\end{figure}

 We consider the density of contact points $g(l,L)$ for various
values of $u$ in the interval where we suspect to have the phase
transition. If we have conformal invariance (see Section 5), we
should have in the finite-size scaling limit
\be \label{6.2}
g(l,L) = C_gL_c^{-\mu},
\ee
where $L_c$ is given by \rf{5.2}, $C_g$ is a constant and $\mu$ a
critical
exponent to be determined. The results of the Monte Carlo simulations
are presented in Fig.~\ref{confinv1} for $L = 4096$. It is shown that for $u =
4.685$ one obtains a very nice fit to the data if one takes:
\be \label{6.3}
g(l,L) = 1.26 L_c^{-1.65}.
\ee

Notice that the value of $\mu$ is very closed to 5/3.

\begin{figure}
\centering
\includegraphics[angle=0,width=0.5\textwidth] {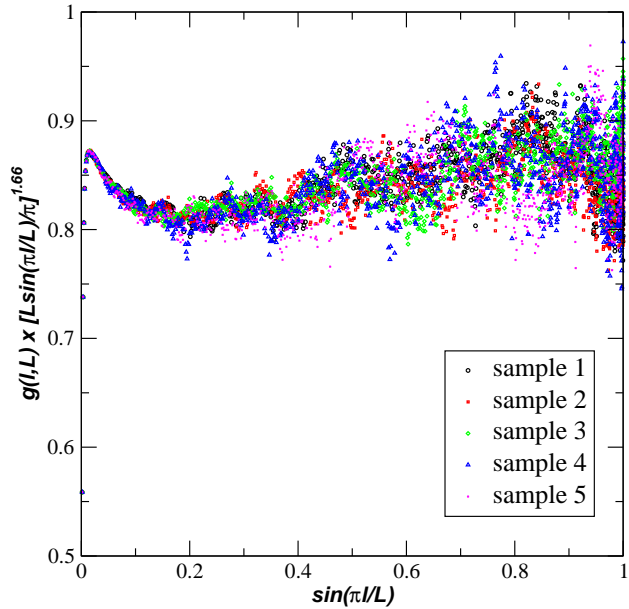}
\caption{
 The density of contact points $g(x,L)$ multiplied by
${(L_c)^{1.66}}$ as a function of ${sin(\pi l/L)}$ for $u = 4.685$ and
$L = 4096$. We show the results of  five samples with $10^9$ Monte Carlo steps}
\label{confinv2}
\end{figure}

\begin{figure}
\centering
\includegraphics[angle=0,width=0.5\textwidth] {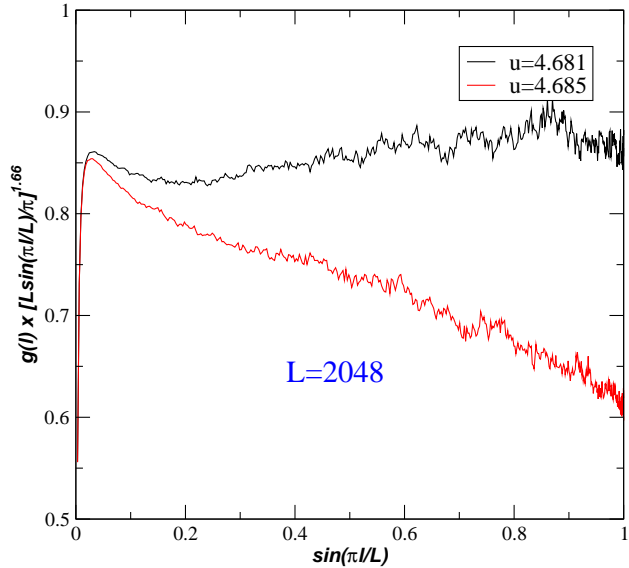}
\caption{
 Effect of the lattice size on the determination of $u_c$ 
for the lattice size $L =
2048$ by averaging $10^{9}$ Monte Carlo steps. The value $u_c = 4.681$ gives a better estimate for
$u_c$ than $u_c = 4.685$ obtained for $L = 4096$.}
\label{confinv3}
\end{figure}

 One can have a different look at the data for $u_c = 4.685$ and $L =
4096$ (see Fig.~\ref{confinv2})  plotting $g(l,L)\times L_c^{1.667}$ as a function of
$\sin(\pi l/L)$. One observes that within the errors, one obtains a
constant, as expected. We have to stress that there is a small drift
in the estimates of $u_c$ with the size of the system. If one uses
the $L = 2048$ data, the value $u_c = 4.681$ is preferred. This
observation is illustrated in Fig.~\ref{confinv3}.

\begin{figure}
\centering
\includegraphics[angle=0,width=0.5\textwidth] {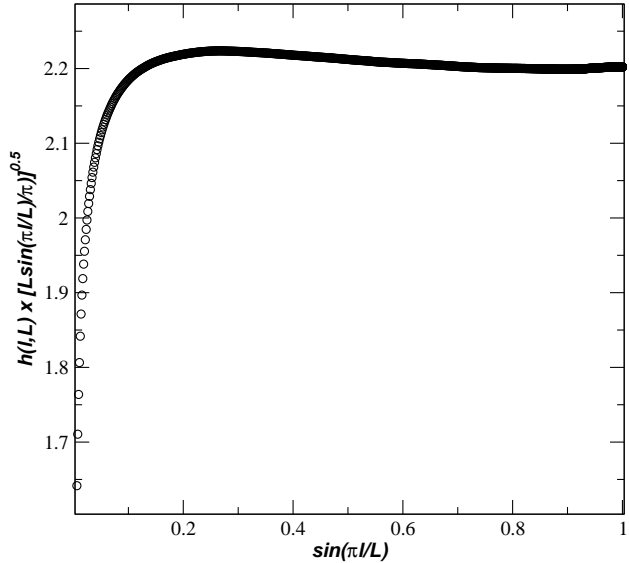}
\caption{
 The average height $h(l,L)$ multiplied by ${L_c}^{1/2}$ as a
function of $\sin(\pi l/L)$ for $u = 4.685$ and $L = 4096$. The 
results shown are obtained by averaging $10^9$ Monte Carlo steps. The estimated 
error are smaller than the symbols in the figure.}
\label{confinv4}
\end{figure}

 The next quantity we are looking at is the average height $h(l,L)$.
If we have conformal invariance, in the finite-size limit we expect
to find the expression:
\be \label{6.4}
h(l,L) = C_h L_c^{\nu},
\ee
where $\nu$ is a critical exponent and $C_h$ is a constant. Using
the data obtained for $u_c= 4.685$ in Fig.~\ref{confinv4} we plot 
$h(l,L)\times{L_c}^{0.5}$
as a function of $\sin(\pi l/L)$ and find almost a constant value. This
implies $\nu = 0.5$. We have measured $\nu$ considering $h(L/2,L)$
for various values of $L$ and our best estimates are  $0.50 \leq \nu \leq 0.52$. An exact estimate  for $\nu$
 is hard to get since, as mentioned above, the estimates of $u_c$
change slightly with $L$.

 Finally one has measured the density of peaks and valleys for
large values of $L$ and obtained
\be \label{6.5}
\tau \approx 0.35.
\ee

 What have we learned up to now about the $u_c$ phase transition? The
data are compatible with conformal invariance. We will also show
below that the critical exponent $z = 1$ and that the spectrum of the
Hamiltonian is what we expect. Although they are both conformal
invariant, the phase transitions at $u = 1$ for the RPM and at $u_c$
for the RSM are different. The average height growth logarithmically in the
RPM and like a power in the RSM. The density of contact points has an
exponent $1/3$ in the RPM and probably $5/3$ in the RSM. Also the density
of peaks and valleys are different. There are fewer peaks and valleys
in the RSM. This can be understood as follows: since the phase
transition takes place at a large value of $u$, in the stationary state, 
many tiles are
adsorbed therefore many have to be desorbed. This implies that one
needs many sites without peaks or valleys.

b) {\it Time dependent phenomena at $u_c$.}

\begin{figure}
\centering
\includegraphics[angle=0,width=0.5\textwidth] {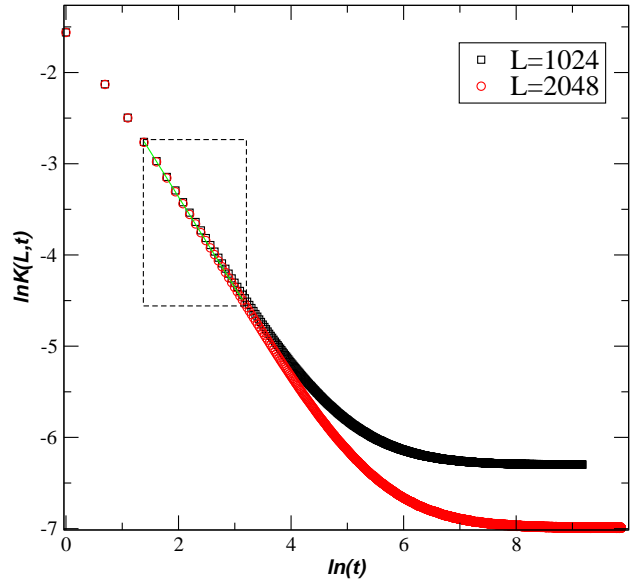}
\caption{
 Determination of the dynamic critical exponent $z$ at $u=4.685$. The
density of clusters versus time on a log scale. 
The substrate was used as the initial condition, and the average was taken over  $10^6$ samples. 
The fitting $y =-1.38 -0.99x$ was 
obtained in the region inside the dashed square. }
\label{confinv5}
\end{figure}

 In order to determine the value of the dynamic critical exponent
$z$, we use the properties of the Family-Vicsek scaling function
\cite{FV}. We first consider the density of clusters $K(L,t)$. From (6.3)
follows that the number of clusters is finite therefore the density
of clusters in the stationary state behaves like
\be \label{6.6}
K(L) = C_K/L,
\ee 
where $C_K$ is a constant. This implies  that for large values of $L$
and short times one expects
\be \label{6.7}
K(t) = \frac{D_K}{t^{1/z}},
\ee
being  $D_K$  a constant. In Fig.~\ref{confinv5} we show the short time
dependence of $K(t)$, and a fit to the data gives
\be \label{6.8}
\ln[K(t)] = -1.38 -0.99\ln(t),
\ee
from which we get the value $z = 1.01$ extremely close to the value
$z = 1$ required by conformal invariance.

\begin{figure}
\centering
\includegraphics[angle=0,width=0.5\textwidth] {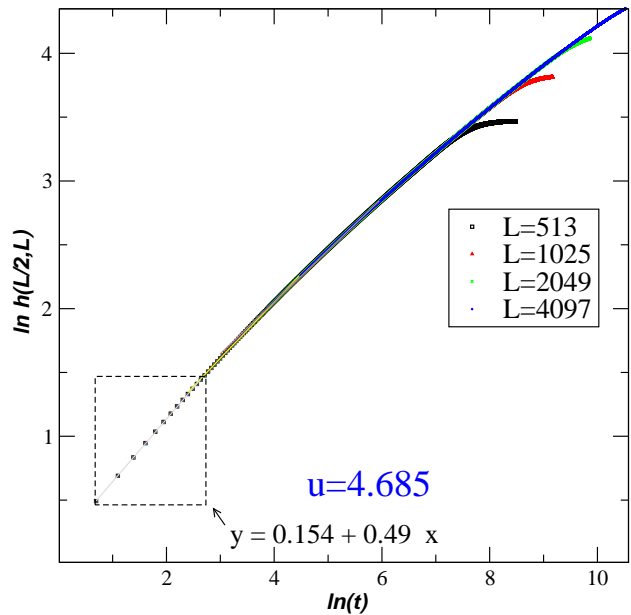}
\caption{
 The average height at $L/2$ as a function of time for
$L = 512, 1024, 2048, 4096$ on log scales. The initial state was the substrate,
 and the average was over $10^6$ samples.
 The region where the fitting was taken is also shown.}
\label{confinv6}
\end{figure}

A similar analysis was done for the average height at the half value
of $L$, $h(L/2,L)$. Since in the stationary state $h(L/2,L) \approx  L^{1/2}$,
we expect, if $z = 1$ that for short times and large values of $L$ to
find $h(L/2, L,t) \approx t^{1/2}$. This expectation is confirmed by the
data shown in Fig.~\ref{confinv6}, where for short times the fit gives:
\be \label{6.9}
\ln [h(L/2,L,t)] = 0.154 + 0.49\ln(t).
\ee

\begin{figure}
\centering
\includegraphics[angle=0,width=0.5\textwidth] {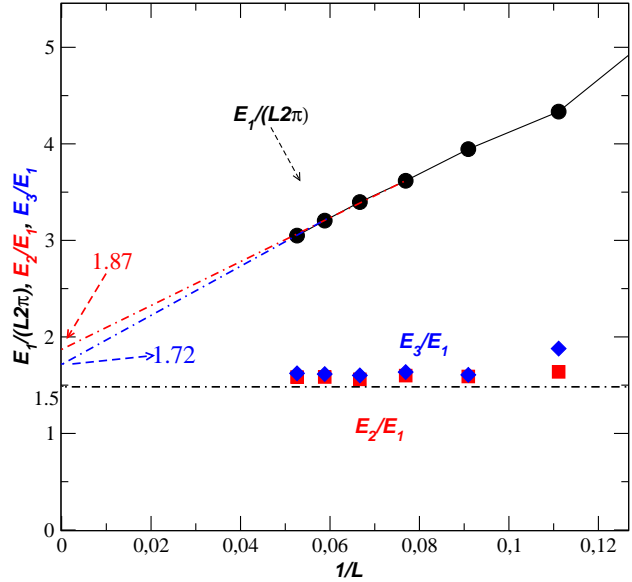}
\caption{
Finite-size scaling analysis of the spectrum of the
Hamiltonian at $u_c = 4.685$. The data are obtained by diagonalizing
the Hamiltonian up to 18 sites. $LE_1(L)/2\pi$, $E_2(L)/E_1(L)$  and
$E_3(L)/E_1(L)$  are shown as a function of $1/L$. The extrapolated 
values for $E_1/(2\pi L)$ is also shown.}
\label{confinv7}
\end{figure}

 A final confirmation of conformal invariance at $u_c$ is obtained
looking at the first excited states of the Hamiltonian \rf{e5p} at $u_c =
4.685$.  We have diagonalized numerically the Hamiltonian up to $L = 18$. In
Fig.~\ref{confinv7} we show for
convenience $L\times E_1(L)/2 \pi$ and 
not  $L\times E_1(L)/\pi$ as a function of
$1/L$. Using the data and (5.7) with $\Delta = 2$ one gets $v_s \approx
3.7$ 
(a value of the same order as the value \rf{5.8} in the RPM).
 We also show the ratios $E_2(L)/E_1(L)$ and $E_3(L)/E_1(L)$, as a function
of $L$. The data nicely converge to the value 3/2 indicating that we
have two levels with $\Delta = 3$ (in the RPM one has only one level
with this value, see (5.10)).

The existence 
of two levels with $\Delta = 3$ is a puzzle. Since the central charge
 of 
the Virasoro algebra is $c = 0$ (for a stochastic 
model the ground state of the Hamiltonian 
is always equal to zero), the character 
corresponding to the vacuum representation is 
$\chi_0(q) = Z(q)$ (see \rf{5.9}). 
There is only one scaling dimension with $\Delta = 3$ (see \rf{5.10}). 
On the other hand, $\Delta = 3$ is not in the Kac table according 
to which the scaling dimensions are
\be \label{6.10}
\Delta_{r,s} = [(3r - 2s)^2 - 1]/24
\ee
with $r$ and $s$ positive integers and therefore one has to add 
to the spectrum a standard 
Virasoro character:
\be \label{6.11}
\chi_3(q) = q^3 \prod_{n=1}^{\infty}(1-q^n)^{-1}.
\ee

 In the finite-size scaling limit, the spectrum of the Hamiltonian is given 
by the 
partition function $\chi_0(q) + \chi_3(q)$. 
This implies the following values of 
$\Delta_i$ and degeneracies ($d_i$):
\be \label{6.12}
\Delta = 0 (1), \quad 2 (1), \quad 3 (2), \quad 4 (3), \quad 5 (4),\quad \ldots .
\ee 

 In order to check if the levels with the corresponding degeneracies 
are seen in our 
model, we have looked at the lower part of the spectrum of the 
Hamiltonian for $L = 18$. 
Normalizing the eigenvalues by taking $E_1(18) = 2$, one gets the 
following values:
\be \label{6.13}
0; \quad 2; \quad 3.16;3.24; \quad 4.35; 4.36; 4.54; \quad 5.20; 5.4; 5.6; 6.0;\ldots
\ee
 We have checked that the values of $E_i(L)/E_1(L)$ converge from 
above from larger values and cluster towards their 
asymptotic ones. We   conclude that, although the lattice sizes are up to 
$L=18$,  at least for the lower part of 
the spectrum there is 
agreement between \rf{6.13} and \rf{6.12}.
 
We have tried to see if at $u_c$, the Hamiltonian has hidden symmetries for
finite $L$. They could show up in the observation of degeneracies of the
spectra at finite $L$. In \rf{6.13} one notices that two couples of values are
very close (3.16 and 3.24 respectively 4.35 and 4.36). Since $u_c$ is not
known exactly we have checked if small changes in the value of $u$ around
the value 4.685 could make the levels degenerate for all values of $L$. Our
investigation gave a negative result. Degeneracies were observed only
for one of the two couples and only for one value of $L$. This observation
probably excludes the existence of Jordan cells in the finite-size scaling
limit since in all known examples the Jordan cells occurring in the
conformal theory appear already for finite values of $L$.

c) {\it The linear growth phase ($u > u_{c}$).}

\begin{figure}
\centering
\includegraphics[angle=0,width=0.5\textwidth] {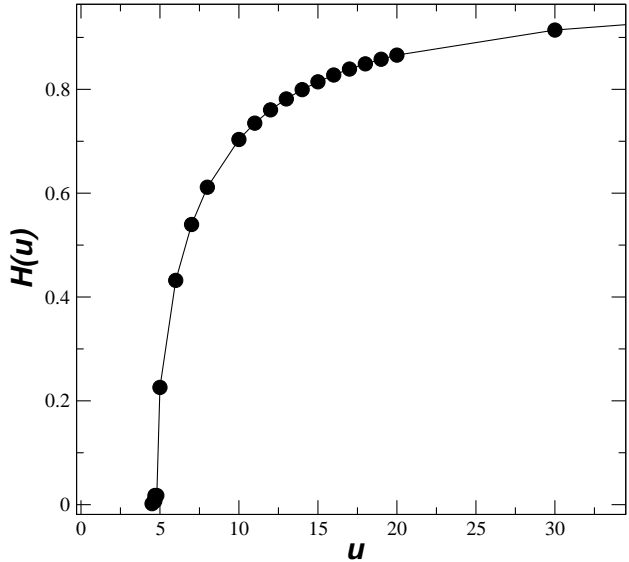}
\caption{
 The $u$ dependence of $H(u)$. Monte Carlo simulations for 
lattice $L = 16384$. The results were obtained by taking $10^7$ Monte Carlo steps. The estimated errors are smaller than the symbols.}
\label{heiglaru}
\end{figure}

 As mentioned earlier, in this phase the 
heights profiles in the stationary state are 
triangles (see Fig.~\ref{rsmhei2} for $L = 2048$) with heights which are fractions of the 
maximum 
possible height $L/2$. The maximum height is seen in the stationary 
states of the FFM. 
We have studied the $u$ dependence of
\be \label{6.14}
H(u) = 2h(L/2,L)/L
\ee
using Monte Carlo simulations for a very large lattice ($L = 16384$). 
The results are shown in Fig.~\ref{heiglaru}. One notices, as expected,  
that $H(u)$ increases with $u$ (for $u$
 infinite, 
$H(u) = 1$, the triangle becomes the pyramid).

 Since $H(u)$ has a clear geometrical meaning, it can be seen as an order 
parameter in the 
LG phase \cite{DER}.

\begin{figure}
\centering
\includegraphics[angle=0,width=0.5\textwidth] {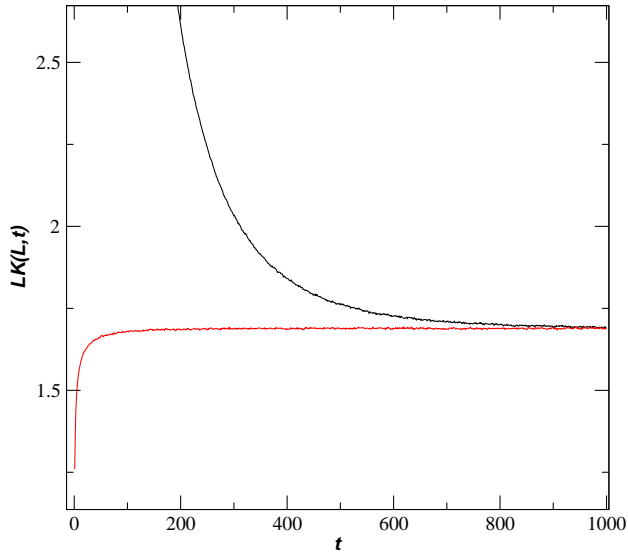}
\caption{
The number of clusters $LK(L,t)$ as a function of time for two
different initial conditions (pyramid in red and substrate in black). The 
lattice size is $L=4096$ and $u=5$. On each evolution there were $4\times 10^5$samples in the Monte Carlo simulations.}
\label{timeevol1}
\end{figure}

 In order to determine the values of the dynamic critical exponent $z$, we 
have performed 
an analysis similar to the one done for $u = u_{c}$ (see eqs.~\rf{6.6} and \rf{6.7}).
 We have 
first convinced ourselves that for $u > 1$ in the stationary state 
and the large $L$ limit 
the number of clusters stays finite. Starting with two different 
configurations (the 
substrate and the pyramid), we have looked at the time 
dependence of the number of 
clusters $C_K (t)$ and checked if they converge at the same value $C_K$. 
The results of the 
computer simulations for $L = 4096$ and $u = 5$ shown in Fig.~\ref{timeevol1} 
 show that this is
 indeed the 
case. This observation implies that in the stationary state, the density of 
clusters   
is given by eq.~\rf{6.6}. The next step is to determine $z$ using eq.~\rf{6.7}. 

\begin{figure}
\centering
\includegraphics[angle=0,width=0.5\textwidth] {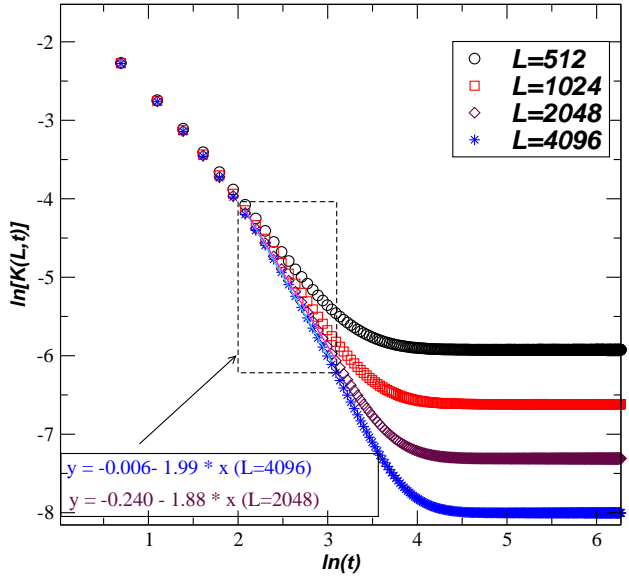}
\caption{Short time dependence of the density of clusters for several lattice sizes of the RSM at $u=7$. It is also shown the fitting results in the region of the dashed 
rectangle for $L=2048$ and $L=4096$. On each evolution there were $4\times 10^5$ samples in the Monte Carlo simulation.} 
\label{timeevol2}
\end{figure}

In Fig.~\ref{timeevol2} we show the short time dependence of the density of clusters for
 various 
lattice sizes and $u = 7$. One has used the substrate as initial condition. Notice the data collapse 
which  allows to determine $z$. A similar analysis was done for other 
values of $u$. Our 
estimates for $z$ are $0.77$ ($u = 5.5$), $0.66$ ($u = 6$), $0.5$ ($u = 7$), $0.43$ ($u = 8$),
 $0.35$ ($u = 10$). 

 We conclude that in  the LG phase one is gapless, the dynamic 
critical exponents 
decrease in value with increasing values of $u$. We had a similar 
behavior in 
the RPM.

\section{ Conclusions}

 We have considered three one-parameter dependent stochastic models
(FFM, RPM and RSM) defined on Dyck paths. Dyck paths have fixed
boundaries \rf{e2a} and are confined to the upper half plane \rf{e2b}.
The parameter denoted by $u$ in the text, is equal to the ratio
of the adsorption and desorption rates. Adsorption is local in the
three models, desorption is local in the FFM and nonlocal in the
other two. The phase diagrams of the three models are similar: for
$u < u_{c}$ the model is gapped, it undergoes a phase transition at 
$u_{c}$ where it is gapless, for $u > u_{c}$ it is again gapped for the FFM 
but gapless for the RPM and RSM in which desorption is nonlocal.

 The aim of the paper was to understand the effect of having
nonlocal rates. The raise and peel model was shortly described 
here because it is, as far as we know, the first intensively 
studied model [5,6] in which the desorption process, which takes
place when a tile hits a slope, is nonlocal. The peaks are not
active. In the raise and strip model presented for the first time in
this paper, the slopes are not active but when a tile hits a peak it
triggers a nonlocal desorption process. In order to 
clarify the role of nonlocality in the latter model, we have
presented the flip-flop model which is similar to the RSM (the peaks
and valleys are the active sites) with the major difference that
desorption is local.

 Two Dyck paths play an important role in these models: the
substrate (no tiles) and the pyramid configuration in which the
height at the middle is equal to half the size of the system.
If $u = 0$, the stationary state is the substrate configuration in all three 
models.

 In the flip-flop model $u_{c} = 1$ and the dynamical critical exponent
$z = 2$ (one has a random walker). We have studied the spectra of the
Hamiltonian for $u < 1$ and $u > 1$ and obtained that the system is 
gapped in both cases. In the stationary state, the average height is
finite for $u < 1$, and increases like $L^{1/2}$ at $u = 1$ ($L$ is the system 
size). For any $u > 1$, in the large $L$ limit, the system has small 
fluctuations around the pyramid configuration 
(the average height at the middle of the system $h(L/2,L)$ is equal
to $L/2$). This result was obtained using known results from 
combinatorics. A variant of this model in which the configuration space is 
changed (RSOS paths without the restriction (2.3)) and the rates are 
local is presented in Appendix A. The results are similar to those 
obtained for the flip-flop model. 

 In the raise and peel model, $u_{c} = 1$, the dynamical critical exponent 
$z = 1$
and one   has conformal invariance. It is this special property which 
inspired our work presented here. Is conformal invariance a consequence of 
nonlocality? 
That this might be the case is suggested by the observation that in the peak 
adjusted raise and peel model \cite{ALCA} where adsorption is also 
nonlocal, one has 
conformal invariance. In the stationary states the average heights are finite for 
$u < 1$ and increase logarithmically with $L$ not only for $u = 1$ but also for 
$u > 1$. There are many properties of this model which are known exactly 
because the Hamiltonian is integrable at $u=u_{c}=1$.
 
 The raise and strip model is most probably not integrable. All our 
results are based on Monte Carlo simulations on large lattices. The best 
estimate for the critical point is $u_{c} = 4.685$. For $u < u_{c}$ and large 
$L$, the average 
values of the heights stay finite  albeit larger than in the 
RPM. For $u = u_{c}$, one obtains $z = 0.99$ very close to the value $z = 1$. 
Several tests suggest that one has conformal invariance. At $u_{c}$ the RSM 
model has different properties than the RPM. The finite-size scaling 
spectrum of the Hamiltonian are given by different Virasoro modules in the 
two models. The critical exponents are also different. The average height 
increases like a power of $L$ and not like $\ln L$. 
For $u > u_{c}$ the dynamical 
critical exponent $z$ decreases if $u$ increases, similar to what was observed 
in the RPM. The analogy stops here however. Whereas in the RPM the average 
height increases logarithmically with $L$, in the RSM the heights profile is 
very different. For large values of $L$, the profile is a triangle with the 
tip increasing linearly with $L$. We have therefore called this domain of $u$, 
the "linear growth phase". For $u$ very large the triangle gets close to the 
pyramid configuration seen in the flip-flop model.

 We believe that the phase diagram observed in the RPM and RSM is general 
for the class of models of nonlocal growth. This has of course to be 
proven looking at other models. Such a study might also bring a better 
understanding of some properties of the RSM (the values of the exponents   
are just one example).          

The relevance of our work for physical systems like polymers with nonlocal 
interactions \cite{p1,p2,p3} remains to be studied.
\section{Acknowledgments}

We would like to thank A. Gainutdinov for suggesting Eq.~\rf{6.12}, 
B. Derrida, H. Hinrichsen and V. B. Priezzhev for discussions. 
   This work was supported in part by FAPESP and CNPq 
(Brazilian Agencies).

\appendix
\section{  Filling a square with tiles}

\begin{figure}
\centering
\includegraphics[angle=0,width=0.25\textwidth] {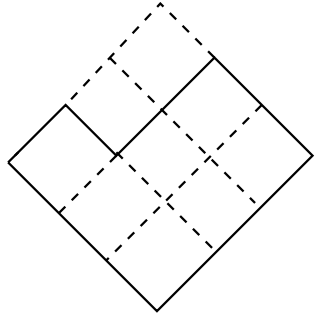}
\caption{
 Configuration space for $L = 6$. The interface is a path which 
separates the seven tiles inside the square from a rarefied gas of tiles.  } 
\label{tile1}
\end{figure}

 This is the flip-flop model described in Section~3 
in the configuration 
space of RSOS paths without the restriction \rf{e2b}.  
The paths are allowed to move in 
the lower half-plane and it is amusing to see what is the effect of 
changing the configuration space. If $u = 1$, one has a random walker with 
fixed ends at positions $0$  and $L$. It is convenient to see the paths as describing an 
interface between tiles which fill part of a square and a rarefied gas 
of tiles. In Figure~\ref{tile1} we show such a path in the case $L = 6$. Out of a maximum
of 9 tiles, the square is filled with 7 tiles.

\begin{figure}
\centering
\includegraphics[angle=0,width=0.65\textwidth] {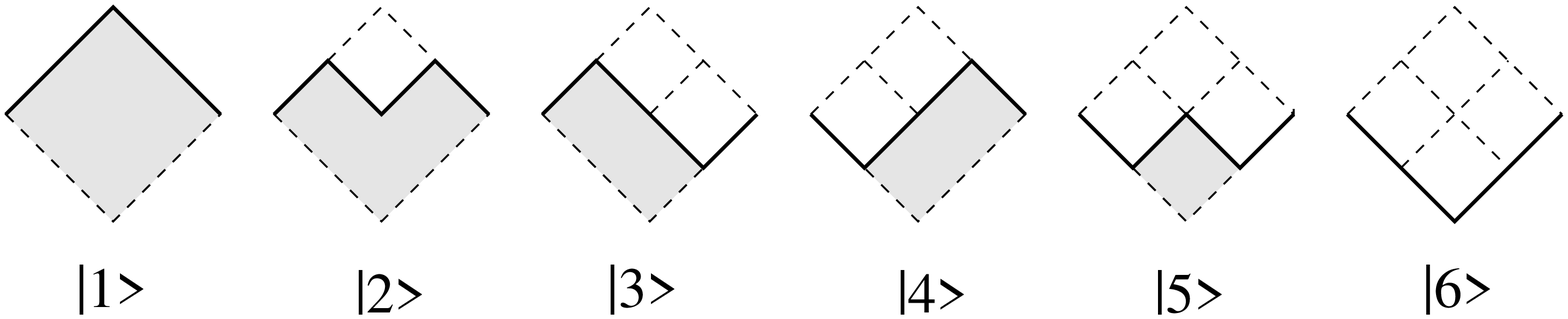}
\caption{  
 The six configurations for $L = 4$.}
\label{tile2}
\end{figure}

 To see how the model works, we take $L = 4$. There are 6 configurations in 
this case as shown in Figure~\ref{tile2}.

 The Hamiltonian obtained using the rules of the flip-flop model is
\ba
 \label{a.1}
&& H =-
\left( \begin{array}{c|rrrrrr}
 & \ket{1} & \ket{2} & \ket{3} & \ket{4} &\ket{5} &\ket{6} \\ \hline
\bra{1} &   -1 &    u &   0  &  0 & 0 & 0\\
\bra{2} &   1 &  -(2+u) &   u & u &  0 &    0 \\
\bra{3} &   0  &   1   & -(1+u) &  0 &    u & 0 \\
\bra{4} & 0  &  1  &  0 &  -(1+u)   &   u & 0  \\
\bra{5} &     0 &  0 &  1 & 1 &  -(2u+1) & u \\
\bra{6} & 0 & 0 &  0 & 0 & 1 & -u
\end{array} \right) .
\ea
The wave function corresponding to the eigenvalue zero is:
\be \label{a.2}
u^4\ket{1} + u^3\ket{2} + u^2(\ket{3} + \ket{4}) +u\ket{5} + \ket{6}.
\ee
In the stationary state, each configuration gets as a coefficient a 
monomial in $u$ with an exponent equal to the number of tiles inside the 
square. The partition function is
\be \label{a.3}
  Z_4(u) = 1 + u + 2u^2 + u^3  + u^4.
\ee 
One could have obtained directly \rf{a.2} using the matrix-product method 
\cite{BEV} for an open system  with the processes given by (4.14) and 
(4.15). If a step up in the 
path is given by a generator $U$ and a step down by a generator $D$, we can use
the algebra
\be \label{a.4}
 UD = u DU
\ee
to find for example, that the coefficient of the configuration $\ket{1}$ ($UUDD$) gets a factor $u^4$ compared with the configuration $\ket{6}$ ($DDUU$).

It is easy to show that the partition function $Z_L(u)$ for $L$ sites ($L = 2n$) is
\be \label{a.5}
Z_{2n}(u) = \frac{(2n)_u!}{((n)_u!)^2} = \frac{(u^{n+1}-1)\cdots (u^{2n}-1)}
{(u-1)(u^2-1)\cdots (u^n-1)},
\ee 
where 
$(n)_u!=2_u3_u\cdots n_u$ are $u$-factorials and
\be \label{a.6}
m_u = \frac{u^m-1}{u-1}
\ee
It is easy to show that the number of tiles inside the square
\be \label{a.7}
N_L(u) = u\frac{d}{du} \ln Z_L(u)
\ee
in the large $L$ limit is:
\ba
\lim_{L} N_L(u) = \left\{
\begin{array}{ll}
n^2 + C(u)& \mbox{if } u>1 \\
n^2/2 & \mbox{if } u=1 \\
-C(1/u) &\mbox{if } u<1 
\end{array} \right.,
\ea
where $C(u)$ is given by 
\rf{4.13}. This result is amusing since it shows 
that for any $u > 1$, in the large $L$ limit, the important configurations are 
the same as for the flip-flop model. For $u < 1$, the relevant configurations
are those near the empty square. They are obtained by a simple reflexion
(top $\rightarrow$ bottom) of the configurations relevant for $u > 1$ (the Hamiltonian 
is invariant under the transformation $u \rightarrow  1/u$ with a change of the time 
scale). At $u = 1$, one has the random walker.

 The phase diagram of this model is the same as the one of the flip-flop 
model. The only difference between the two models in the large $L$ limit is 
the heights profiles in the $u < 1$ domain.

\end{document}